\documentclass[12pt,a4paper,onecolumn,nofootinbib]{revtex4}
\usepackage{amsmath}
\usepackage{amssymb}
\usepackage{epsf}
\usepackage{graphicx}
\usepackage{color}

\newcommand{\rot}{\mathop{\mathrm{rot}}\nolimits}

\renewcommand{\Im}{\mathop{\mathrm{Im}}\nolimits}

\newcommand{\divergency}{\mathop{\mathrm{div}}\nolimits}

\graphicspath{{img/}}

\begin{document}

\title{The electron beam instability in a one-dimensional cylindrical photonic crystal}

\author{\firstname{V.G.} \surname{Baryshevsky}}
\email[]{bar@inp.minsk.by}
\affiliation{Research Institute for Nuclear Problems of Belarussian State University,
Bobruiskaya Str.~11, 220050 Minsk,
Belarus}

\author{\firstname{E.A.} \surname{Gurnevich}}
\email[]{genichgurn@gmail.com}
\affiliation{Research Institute for Nuclear Problems of Belarussian State University,
Bobruiskaya Str.~11, 220050 Minsk,
Belarus}

\begin{abstract}
The radiative instability of the relativistic electron beam in a
periodic dielectric-filled cylindrical waveguide is considered.
The dependence of the beam instability increment on the radiated
wave frequency near the region of dispersion equation roots
degeneracy is studied. It is shown that sharp change in the
instability of the beam under the conditions of two-wave
diffraction in Compton generation regime, making the increment
proportional to the fourth root from the beam density $\rho^{1/4}$
in contrast to conventional law $ \rho^{1/3}$, brings radiation
generation (amplification) in the considered system to be
essentially improved in comparison with conventional devices (BWO,
TWT, FEL etc). Numerical calculations of the instability increment
for various parameters of the system are performed.
\end{abstract}

\maketitle

\section{Introduction}
Currently, there is a large number of generators and amplifiers of
electromagnetic radiation (from microwave to optical wavelengths
range) based on electron beams, for example traveling-wave tubes
(TWT), backward wave oscillators (BWO), free electron lasers
(FEL), ubitrons etc  \cite{Trubeckov, Silin1971}. It is known that
any radiating system is characterized by its dispersion equation
describing in the case of small perturbations the possible types
of waves in the system. A detailed analysis of the properties of
this dispersion equation \cite {Gover} shows that the gain in the
Compton regime (increment of electron beam instability) of the
most commonly used generators (TWT's, BWO's, FEL's) is
proportional to $\rho^\frac{1}{3}$, where $\rho$ is the density of
the electron beam. However, in the papers \cite
{Baryshevsky1984,Baryshevsky1988} it is found that for electron
beam moving in spatially periodic medium under the conditions
providing the coincidence of roots of the dispersion equation
there is a new physical law: the increment of instability is
proportional to $\rho^{1/(3 + s)}$, where $s$ is the number of
additional waves appearing due to diffraction in the crystal. The
analysis shows that this new law leads to reducing of the electron
beam current density,
 needed to reach lasing threshold. This enables development of a new type of
Free Electron Lasers called
the Volume Free Electron Lasers (VFEL) \cite{Baryshevsky1988,Baryshevsky1, Baryshevsky2002}.
 Due to a significant change in the threshold
conditions, VFEL can provide a more efficient radiation process than conventional generators.

Until recently the theoretical studies of the problem of beam instability in
photonic crystals were carried out for the case of infinite in the transverse direction
crystals. However, in many cases the mode structure of the electromagnetic field in laser cavities
can not be neglected (for example, when the lasing is performed in the microwave range).
For the first time the process of VFEL lasing in photonic crystal finite in transverse direction
is considered in \cite{Baryshevsky2010} . In \cite{Baryshevsky2010} it is also pointed out that the four-fold
degeneracy
of the dispersion equation roots (when increment is
$\sim\rho^{1/4}$) is possible in the one-dimensional photonic crystal when the condition
 $\varepsilon_0>1$ (in the conventional generators - BWO, TWT etc. $\varepsilon_0$ is usually  equal to
$1$) is fullfilled. In this paper we consider the simplest
example of such one-dimensional photonic crystal  -- a cylindrical waveguide with a periodic
dielectric filling.

\section{Basic formulas}
The system of equations describing the interaction of an electron beam with an electromagnetic wave in a waveguide
 can be obtained from the Maxwell and electron movement equations. We rewrite the Maxwell equations as
follows
\begin{equation}
 \rot\rot \vec{E}(\vec{r},\omega) - \frac{\omega^2}{c^2}\varepsilon(\vec{r},\omega)
\vec{E}(\vec{r},\omega) = \frac{4\pi i \omega}{c^2}\vec{j}(\vec{r},\omega),
\label{eq:maks}
\end{equation}
\begin{equation}
 \divergency \varepsilon (\vec{r},\omega)\vec{E}(\vec{r},\omega) =
4\pi\rho(\vec{r},\omega),
\label{eq:divergency}
\end{equation}
\begin{equation}
 i\omega\rho(\vec{r},\omega)-\divergency \vec{j}(\vec{r},\omega)=0,
 \label{eq:continuity}
\end{equation}
where $\vec{E}(\vec{r},\omega)=\int\vec{E}(\vec{r},t)e^{i\omega t}dt$ is the Fourier transformation of the
electric field $\vec{E}(\vec{r},t)$; $\varepsilon(\vec{r},\omega)$ is the dielectric permittivity of the medium
filling waveguide; $\vec{j}(\vec{r},\omega)$ and $\rho(\vec{r},\omega)$ are Fourier transformations of the
current
density $\vec{j}(\vec{r},t)$ and electric charge density  of the beam $\rho(\vec{r},t)$, respectively. The
quantities $\vec{j}(\vec{r},t)$ and $\rho(\vec{r},t)$ can be expressed in following way
\begin{equation}
 \vec{j}(\vec{r},t)=e\sum_\alpha\vec{v}_\alpha(t)\delta(\vec{r}-\vec{r}_\alpha(t)),
 \label{eq:j_rt}
\end{equation}
\begin{equation}
 \rho(\vec{r},t)=e\sum_\alpha\delta(\vec{r}-\vec{r}_\alpha(t)),
 \label{eq:rho_rt}
\end{equation}
where $\vec{r}_\alpha(t)$, $\vec{v}_\alpha(t)$ are $\alpha$th electron radius-vector and velocity; the sum in
\eqref{eq:j_rt}-\eqref{eq:rho_rt} is over all electrons in the beam.

The electron movement equations can be written in the form
\begin{equation}
 \frac{d\vec{v}_\alpha(t)}{dt}=\frac{e}{m\gamma}\left\{ \vec{E}(\vec{r}_\alpha(t),
t) + \frac{1}{c} \left[ \vec{v}_\alpha(t)\times \vec{H}(\vec{r}_\alpha(t),t)
\right] -
\frac{\vec{v}_\alpha}{c^2}\left(\vec{v}_\alpha\vec{E}(\vec{r}_\alpha(t),t)\right)
\right\},
\label{eq:movement}
\end{equation}
where $\vec{E}(\vec{r}_\alpha(t),t)$ and $\vec{H}(\vec{r}_\alpha(t),t)$ are
the electric and magnetic field of the electromagnetic wave in the point $\vec{r}_\alpha(t)$ at the time moment $t$,
$\gamma=(1-\frac{v_\alpha^2}{c^2})^{-\frac{1}{2}}$.

Let the $z$ coordinate axis coincide with waveguide axis. We also suppose that electron beam is ``cold''
(velocity spread of the electrons can be neglected) and initial electrons velocities are directed
along $z$-axis ($\vec{u}=u\vec{e}_z$; $(\vec{e}_x,\vec{e}_y,\vec{e}_z)$ are unit vectors of coordinate system). The
dielectric permittivity inside the waveguide is $\varepsilon(\vec{r},\omega) = \varepsilon_0(\vec{\rho}) +
\chi(\vec{r},\omega)$, where $\chi$ is periodic function of $z$: $\chi(\vec{r},\omega) = \sum\limits_{\tau \neq 0}
\chi_\tau(\vec{\rho})e^{i\tau z}$; $\chi_0\equiv\varepsilon_0-1$. With the help of
(\ref{eq:divergency},\ref{eq:continuity}) and supposing that $|\chi(\vec{r},\omega)|\ll 1$  we can rewrite
\eqref{eq:maks} in the following way
\begin{equation}
 -\vec{\nabla}^2\vec{E}(\vec{r},\omega) -
\frac{\omega^2}{c^2}\varepsilon(\vec{r},\omega)\vec{E}(\vec{r},\omega)\approx
\frac{4\pi i}{c^2}\left( \vec{j}(\vec{r},\omega) + \frac{c^2}{\omega^2
\varepsilon_0} \vec{\nabla}(\vec{\nabla}\vec{j}(\vec{r},\omega))\right).
\label{eq:main}
\end{equation}
The field $\vec{E}(\vec{r},\omega)$ can be decomposed in terms of the waveguide eigenfunctions
\begin{equation}
 \vec{E}(\vec{r},\omega) =
\frac{1}{2\pi}\sum\limits_n \int a_n(k_z)\vec{Y}_n (\vec{\rho},k_z) e^{ik_zz}dk_z,
\label{eq:field_in_waveguide}
\end{equation}
where $a_n(k_z)$ are the expansion coeffitients (amplitudes), $\vec{Y}_n(\vec{\rho},k_z)$ and $\kappa_n$ are the
waveguide eigenfunctions and corresponding eigenvalues \cite{Landau, Jackson1965, Mors1953}.

The beam current appearing on the right-hand side of \eqref{eq:main} is a
complicated function of the field $\vec{E}$. To study the problem of the system
instability, it is sufficient to consider the system in the approximation linear
over perturbation, i.e., one can expand the expressions for $\vec{j}(\vec{r},\omega)$ over the
field amplitude  $\vec{E}(\vec{r},\omega)$:
$\vec{j}=\vec{j}_0+\delta\vec{j}$, where $\vec{j}_0$ is the beam current not perturbated by the radiated field,
$\delta\vec{j}\sim \vec{E}(\vec{r},\omega)$ is the beam current induced by the radiated field.
With the help of \eqref{eq:j_rt} one can find
\begin{equation}
 \delta\vec{j}(\vec{k},\omega)= e\sum_\alpha e^{-i\vec{k}\vec{r}_{\alpha
0}} \left\{ \delta\vec{v}_\alpha(\omega-\vec{k}\vec{u}) + \frac{1}{\omega-
\vec{k}\vec{u}}\vec{u}\left(\vec{k}\delta\vec{v}_\alpha(\omega-\vec{k}\vec{u}
)\right)\right\}.
\label{eq:delta_j}
\end{equation}
The quantity $\delta\vec{v}_\alpha(\omega)$ can be obtained from movement equations:
\begin{multline}
  \delta\vec{v}_\alpha(\omega)=\\=\frac{ie}{m\omega \gamma}\int_{}^{}\frac{d^3k'}{(2\pi)^3}e^{i\vec{k}'\vec{r}_{\alpha
0}}\left\{
  \frac{\omega}{\omega + \vec{k}'\vec{u}}\vec{E}(\vec{k}',\omega + \vec{k}'\vec{u}) + \left(\frac{\vec{k}'}{\omega +
\vec{k}'\vec{u}}-\frac{\vec{u}}{c^2}\right)\cdot \left(\vec{u}\vec{E}(\vec{k}',\omega+\vec{k}'\vec{u})\right)\right\}.
  \label{eq:delta_v}
\end{multline}
After substituting \eqref{eq:delta_v} into \eqref{eq:delta_j} in the expression for the current density appear
the sum $\sum\limits_\alpha e^{i(\vec{k}'-\vec{k})\vec{r}_{\alpha 0}}$. Let us average this sum over
distribution of the particles in the beam
\begin{equation}
  \sum\limits_\alpha e^{i(\vec{k}'-\vec{k})\vec{r}_{\alpha0}} \simeq \Phi(\vec{k}_\perp - \vec{k}'_\perp)
  \cdot (2\pi)^3 n_0 \delta(k_z-k_z'),
  \label{eq:phi}
\end{equation}
where $\Phi(\vec{k}_\perp-\vec{k}'_\perp)=\frac{1}{(2\pi)^2}\int_{S}e^{-i(\vec{k}_\perp-
\vec{k}'_\perp)\vec{\rho}}\varphi(\vec{\rho})d^2\vec{\rho}$,
$\frac{1}{S}\int_{S}^{}\varphi(\vec{\rho})d^2\vec{\rho}=1$, $S$ is cross-section of the waveguide,
$n_0$ is the electron density of the beam, $k_z$ is the longitudinal component of the wave vector, the function
$\varphi(\vec{\rho})$ describes the distribution of the
 electrons in the beam cross-section.

The expressions \eqref{eq:main}-\eqref{eq:phi} allow us to write the following system of equations
for amplitudes $a_m(k_z)$ (this system is similar to that describing the multiwave dynamical diffraction
in crystals):
\begin{multline}
  \left(k_z^2-\left(\frac{\omega^2}{c^2}\varepsilon_0-\kappa_m^2\right)\right)a_m(k_z) -
  \frac{\omega^2}{c^2}\sum_m\sum_{\tau\neq0}\chi_{eff}^{mn}(k_z,k_z-\tau)a_n(k_z-\tau)=\\=
  -\frac{\omega_l^2}{\gamma}\frac{1}{(\omega-k_zu)^2}\sum_nA_{mn}a_n(k_z),
  \label{eq:system}
\end{multline}
where $\omega_l^2 = \frac{4\pi e^2 n_0}{m}$ is the Langmuir frequency of the beam, effective susceptibility
$\chi_{eff}^{mn}$ is
\begin{equation}
  \chi_{eff}^{mn}(k_z,k_z-\tau)=\int \vec{Y}_m^*(\vec{\rho},k_z)\chi_\tau\vec{Y}_n(\vec{\rho},k_z-\tau)d^2\vec{\rho}.
  \label{eq:chi_effective}
\end{equation}
Coeffitients $A_{mn}$ in \eqref{eq:system} are
\begin{multline}
  A_{mn} = \frac{1}{c^4}\int \frac{d^2\vec{k}_\perp}{(2\pi)^2}d^2\vec{\rho} d^2\vec{\rho}'
  \varphi(\vec{\rho}') e^{i \vec{k}_\perp (\vec{\rho}-\vec{\rho}')} \times \\ \times
  \vec{Y}_m^*(\vec{\rho},k_z)
  \left(\vec{u} - \frac{c^2}{\omega \varepsilon_0}\vec{k}\right)
  \left(-ic^2\vec{k}_\perp \vec{\nabla}'_\perp + \frac{c^2-u^2}{u^2}\omega^2 \right)
  \left( \vec{u}\vec{Y}_n(\vec{\rho}',k_z) \right).
  \label{eq:ann_coefficients}
\end{multline}

\section{Increment of electron beam instability}
Let us consider the generation on the most commonly used $E_{01}$-mode of circular waveguide.
The vector eigenfunction $\vec{Y}_1(\vec{\rho},k_z)$ corresponding to $E_{01}$-mode can be written as \cite{Mors1953}
\begin{equation}
  \vec{Y}_1 = \frac{1}{\sqrt{\pi}RJ_1(\nu_{01})}\left\{\frac{-ik_z}{\sqrt{k_z^2+
  \kappa^2}}J_1(\rho\kappa)\vec{e}_r + \frac{\kappa}{\sqrt{k_z^2 + \kappa^2}}J_0(\rho\kappa)\vec{e}_z
  \right\},
  \label{eq:eigen_E01}
\end{equation}
where $\nu_{01} \approx 2.4048$, $\kappa=\nu_{01}/R$, $R$ is the waveguide radius, $J_0$ and $J_1$ are Bessel functions.
The effective susceptibility
according to \eqref{eq:chi_effective} is
\begin{equation}
  \chi_{eff}(k_z,\tau) = \chi_\tau\int\vec{Y}_1^*(\vec{\rho},k_z)\vec{Y}_1(\vec{\rho},k_z-\tau)d^2\vec{\rho}
  =\chi_\tau\left(1-\frac{k_z\tau}{k_z^2+\kappa^2}\right).
  \label{eq:effective_permittivity_E01}
\end{equation}
The equations \eqref{eq:system} for the $E_{01}$-mode can be written as
\begin{equation}
  (\omega-k_zu)^2\left( (k_z^2-k_{z0}^2)a(k_z) -k_0^2\sum\limits_{\tau\neq0}\chi_{eff}(k_z,\tau)a(k_z-\tau) \right)
  = \frac{\omega_l^2}{\gamma}A_{11}a(k_z),
  \label{eq:E01_with_beam}
\end{equation}
where $k_{z0}=\sqrt{\frac{\omega^2}{c^2}\varepsilon_0-\kappa^2}$ and $k_0=\omega/c$. We
assume that unperturbed beam
occupies the entire cross section of the waveguide and $\varphi(\vec{\rho})=1$.
 Then the calculation of the $A_{11}$  in
accordance with \eqref{eq:ann_coefficients} gives
\begin{equation}
  A_{11}=\left(\frac{k_0^2}{\gamma^2}+\beta^2\kappa^2\right)\frac{\kappa^2}{k_z^2+\kappa^2}.
  \label{eq:A_coeff_E01}
\end{equation}

Let us consider firstly the case of waveguide with homogeneous filling ($\chi_\tau=0$). Apparently
in this case
the diffracted wave is absent and system \eqref{eq:E01_with_beam}  is reduced to a single equation.
Suppose that right-hand side of \eqref{eq:E01_with_beam} is small. Since the nonlinearity is insignificant, let us
consider as the zero approximation the spectrum of the waves of equation \eqref{eq:E01_with_beam} with zero right-hand
side. So, the solution of \eqref{eq:E01_with_beam} is $k_{z1}=k_{z1}'+\delta k_{z1}$, where $\delta k_{z1}$ is to be
found, and $k_{z1}'$ is the solution of  \eqref{eq:E01_with_beam} with zero right-hand side
$k_{z1}'=k_{z0}=\sqrt{\frac{\omega^2}{c^2}\varepsilon_0-\kappa^2}$. Repeating exactly the arguments, for example
\cite{Baryshevsky2010},
we arrive the following expression for increment
\begin{equation}
  \Im k_{z1} = \Im \delta k_{z1} = \frac{\sqrt{3}}{2}\left(\frac{\omega_l^2}{2 k_{z1}' u^2 \gamma}A_{11}\right)^{
  \frac{1}{3}} \sim \rho^{1/3}.
  \label{eq:increment_one_wave}
\end{equation}

Now let us consider the case $\chi_\tau\neq 0$, $\chi_\tau\ll1$. Apparently when the diffraction conditions are
 not fullfilled we get the same expression \eqref{eq:increment_one_wave} for increment. We therefore assume that the
conditions of two-wave dynamical diffraction in a waveguide are realized, i.e. the wave amplitude $a(k_z+\tau)$
is comparable with the amplitude $a(k_z)$. In this case the dispersion equation has the next form:
\begin{equation}
  (\omega - k_zu)^2 \left\{(k_z^2-k_{z0}^2)(k_{z\tau}^2-k_{z0}^2)-k_0^4\chi_{eff}(k_{z\tau},\tau)\chi_{eff}(k_z,-\tau)
  \right\}=-\frac{\omega_l^2}{\gamma}A_{11}(k_{z\tau}^2-k_{z0}^2).
  \label{eq:disp_beam_two_wave}
\end{equation}
The solution of this equation is $k_{z2}=k_{z2}'+\delta k_{z2}$, where
\begin{equation}
 k_{z2}'^{(1,2)} = k_{z0}\left\{1 - \frac{1}{4}\alpha_B\beta \pm \sqrt{(\alpha_B\beta)^2 +
4\frac{r}{\gamma_0^4}\beta}\right\},
 \label{eq:cold_roots}
\end{equation}
$r=\chi_{eff}(k_{z0}+\tau,\tau)\chi_{eff}(k_{z0},-\tau)$,
$k_{z0}=\sqrt{\frac{\omega^2}{c^2}\varepsilon_0-\kappa^2}$, $\gamma_0=\frac{k_{z0}}{\omega/c}$,
$\beta=\frac{k_{z0}}{k_{z0}+\tau}$ is the diffraction asymmetry factor,
$\alpha_B=\frac{(2k_{z0}+\tau)\tau}{k_{z0}^2}$ is the off-Bragg parameter ($\alpha_B=0$ when the
Bragg condition of diffraction is exactly fulfilled), $\tau=2\pi/D$, $D$ is the
waveguide period. The calculation gives the following expression for increment $\Im k_z$ in the vicinity of the roots
degeneracy point
\begin{equation}
  \Im k_{z2} = \Im \delta k_{z2} \approx
  \left(\frac{\omega_l^2 k_0^2 \sqrt{r}}{\gamma u^2 \tau^2}A_{11}\right)^{1/4} \sim \rho^{1/4}.
  \label{eq:increment_two_wave}
\end{equation}
From \eqref{eq:increment_one_wave} and \eqref{eq:increment_two_wave} we have
\begin{equation}
  \frac{\Im k_{z2}}{\Im k_{z1}}\approx \left(\frac{\omega_l^2}{\omega^2}\frac{\tau^2}{k_0^2}
  \frac{A_{11}}{k_0^2}\frac{1}{(\sqrt{r})^3\beta^2 \gamma}\right)^{-1/12}\gg 1,
  \label{eq:increments_comparation}
\end{equation}
because $\omega_l^2\ll \omega^2$, $A_{11}\ll k_0^2$. This means under diffraction conditions the gain increases.
As a result, the threshold current density in two-wave diffraction case is lower ($j\sim
\frac{1}{(kL)^3(k\chi_\tau L)^{2s}}$, where $L$ is the interaction length).

\section{Numerical solutions}
To be able to experimentally observe this effect it is necessary to know
how accurately the diffraction condition must be satisfied. To study this, the equations
\eqref{eq:disp_beam_two_wave} were solved numerically.
Now we demonstrate the results of this calculation by the example of one of the systems studied
($R=6$~cm, $D=3.6$~cm, $\varepsilon_0=1.23$, $\chi_\tau=0.05$).

\begin{figure}[htp!]
  \begin{center}
  \includegraphics{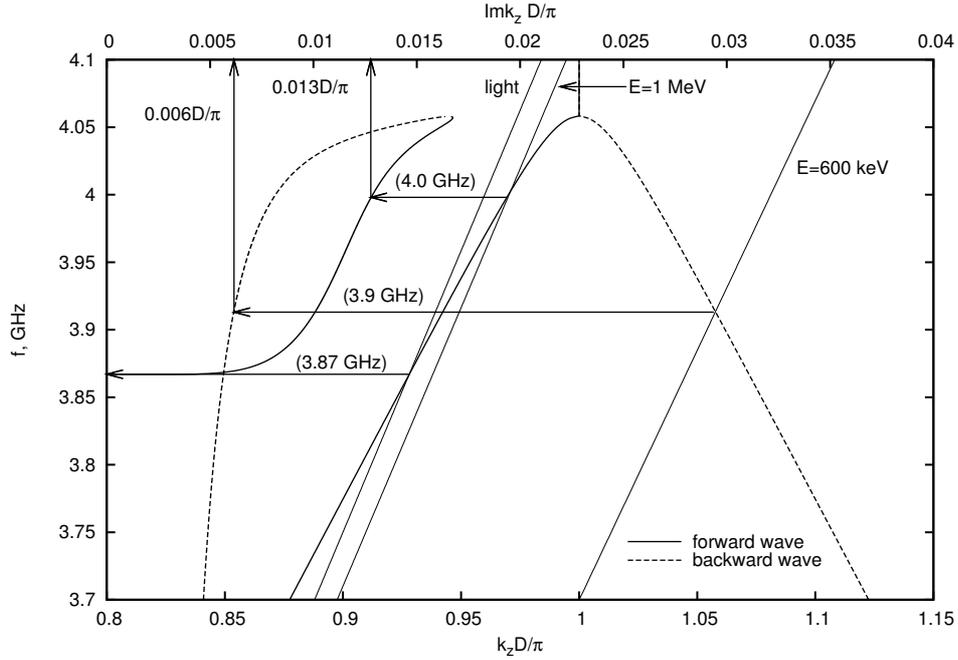}
  \end{center}
  \caption{The dependence of instability increment of electron beam on the frequency near the point of the
roots degeneracy of dispersion equation. Beam current is 0.1~kA. Solid and dashed curves on the right-hand side
of the graph are dispersion characteristics of the forward and backward waves in the waveguide, respectively;
on the left-hand side are plotted corresponding increments. On the graph are also shown the beam lines
($\omega=k_zu$) for the energies 600~keV and 1~MeV and the light line ($\omega=k_zc$).}
  \label{fig:increm_basic}
\end{figure}

\begin{figure}[htp!]
  \begin{center}
    \includegraphics{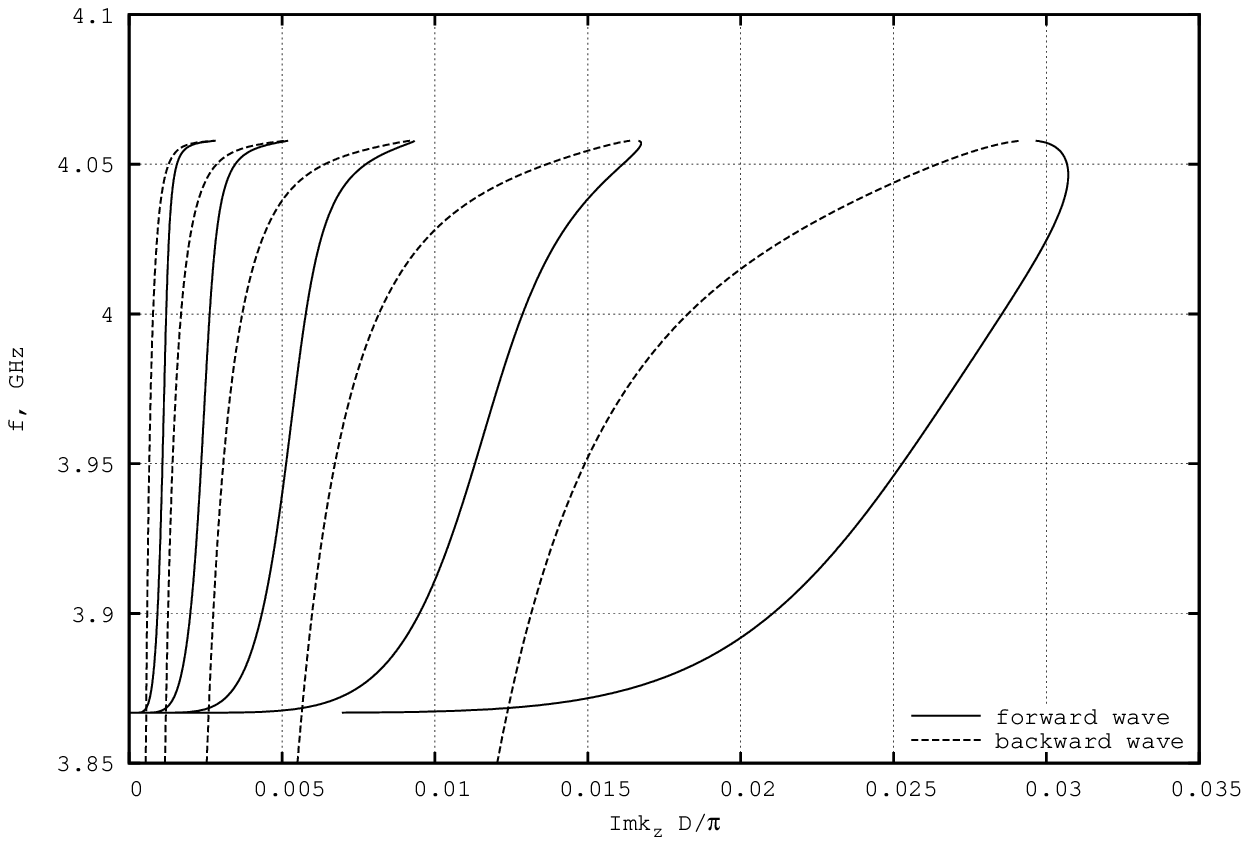}
  \end{center}
  \caption{The growth of the increment with beam current increase. On the graph are shown the increments for cases when
the beam current is 0.1~A, 1~A, 10~A, 0.1~kA and 1~kA.}
  \label{fig:inc_comp1}
\end{figure}

On the fig.~\ref{fig:increm_basic} one can see the dependence of the increment of
instability on the frequency
near region of dispersion roots degeneracy ($4.06$~GHz in our case). Let us explain how the curves in
fig.~\ref{fig:increm_basic} are obtained. Initially, we
found the solutions of dispersion equation without beam \eqref{eq:cold_roots}.
They are shown on the right-hand side of the graph. Next, for each
frequency $f_i$ from the selected range the set of equations
\eqref{eq:disp_beam_two_wave} was solved and the imaginary part of the
solution found $\Im k_z$ was plotted. In the calculations we assumed that the beam current $I$ is given;
the beam energy $E$ was chosen so that the Cherenkov synchronism occurred at a frequency $f_i$.
 Calculations were performed for the two roots of the dispersion
equation.\footnote{Far from the roots degeneracy point they correspond to the so-called forward and backward waves in
the waveguide. But near the point of degeneracy such division, generally speaking, is incorrect because in this case
``forward'' and ``backward'' waves are coupled. Nevertheless, in the article we use the established
terms ``forward'' and ``backward'' waves to refer to the roots of the dispersion equation, since we investigate the
behavior of the system over a wide frequency range (far from the point of degeneration and in the immediate vicinity
from it).
}
For example, for
a backward wave synchronism with the beam with the energy $600$~keV occurs at a frequency of $\sim3.9$~GHz;
if we choose the beam current equal to $0.1$~kA, the corresponding increment will be
$\Im k_z\approx 0.006\pi/D\approx 0.007$~cm$^{-1}$, which is shown in Fig.~\ref{fig:increm_basic}.
A similar example is given on the fig.~\ref{fig:increm_basic} for the forward wave.

From Fig.~\ref{fig:increm_basic} we can trace back the typical dependence of the
increment value on the frequency near the point of the roots degeneracy for the
systems studied. It is easy to see that the synchronism condition for the forward wave
is possible only for frequencies above $f_0$ (in our case $f_0\approx3.87 $~GHz), because velocity of the beam can not
exceed the speed of
light in vacuum $c$. Therefore, for the forward wave the increment begins to rise from
zero at the frequency $f_0$ to a maximum which can be observed both
at the point of the roots degeneracy, and
at some distance from it (see fig.~\ref{fig:inc_comp1}). It is seen that in the vicinity of the degeneracy point the
increment increases with
the
frequency faster than away from the diffraction conditions.
For the backward wave a similar but much more pronounced picture can be seen:
relatively slow increment growth away from the diffraction conditions is replaced by the rapid growth near the
degeneracy point. In addition, for backward-wave the increment has the highest value exactly in the
degeneracy point.

Fig.~\ref{fig:inc_comp1} shows the increment of instability for different beam currents.
For the forward wave the increase of the beam current shifts down the frequency at which
the increment reachs maximum. When the current rises, the
increment maximum becomes wider and lower. Nevertheless, for the
chosen parameters of the waveguide the gain for a backward wave in the vicinity of the degeneracy point
can be several times (from 2 to 5) greater than that for the usual no-diffraction case. But to use this in
practice the synchronism conditions must be satisfied very precisely. For example, when the beam current
is $1$~kA as shown in fig.~\ref{fig:inc_comp1} the imaginary part $\Im k_z$ is reduced by half
when the displacement from the point of degeneracy (4.06~GHz) down in frequency is only $\sim 116 $~MHz
(the peak FWHM is $\sim 116$~MHz). In practice, the control of oscillation frequency with such
precision is not always possible due to many factors, such as the presence of an electron velocity
spread in the beam. Nevertheless, one can find conditions (such as geometric dimensions of the system, values of
$\varepsilon_0$, $\chi_\tau$, the beam energy, etc.) under which it will be possible to perform
the lasing near degeneracy point with sufficient
accuracy. In particular, by use of fig.~\ref{fig:increm_basic} it is easy to find that for selected
geometry the spread in the  electron velocities does not play a significant role due to high electron beam energy
(for the detection of the effect is enough to have $\delta E\leq$ 30\% at the energy $E\sim 1.6$~MeV).

\section{Conclusion}

We conclude with a remark on the above mentioned model of the
one-dimensional crystal. This model is is somewhat idealized since
in real systems it is impossible to use the waveguides with a
solid dielectric filling. As a rule, in practice are applied
waveguides with inner dielectric liner \cite{Peter1992} (partially
filled waveguides). However in considering of the partially filled
waveguide all the above arguments remain valid. In fact, you only
need to substitute in the above model other eigenfunctions and
derive the expressions for coefficients $\chi_{eff}$ and $A_{mn}$.
But the behavior of the instability increment near the degeneracy
points remains the same. Moreover to estimate the increment value
you can use the model of a waveguide with a solid filling by
introducing the average over the cross section of the waveguide
dielectric permittivity $\varepsilon(\omega)$ and taking into
account the fact that with a wave interacts effectively only part
of the beam (located at distance of
$\lesssim\lambda\beta\gamma/(4\pi)$ from the surface of the
dielectric).


It is shown that for a one-dimensional photonic crystal the gain
near region of dispersion equation roots degeneracy rapidly
(several times) increases. With a reasonable choice of the
waveguide (crystal) and electron beam parameters (in particular,
at sufficiently high beam energy) this effect is experimentally
observable and can be used to improve the characteristics of
generators and amplifiers of microwave radiation.

\bibliography{npcs}

\begin{thebibliography}{12}
\expandafter\ifx\csname natexlab\endcsname\relax\def\natexlab#1{#1}\fi
\expandafter\ifx\csname bibnamefont\endcsname\relax
  \def\bibnamefont#1{#1}\fi
\expandafter\ifx\csname bibfnamefont\endcsname\relax
  \def\bibfnamefont#1{#1}\fi
\expandafter\ifx\csname citenamefont\endcsname\relax
  \def\citenamefont#1{#1}\fi
\expandafter\ifx\csname url\endcsname\relax
  \def\url#1{\texttt{#1}}\fi
\expandafter\ifx\csname urlprefix\endcsname\relax\def\urlprefix{URL }\fi
\providecommand{\bibinfo}[2]{#2}
\providecommand{\eprint}[2][]{\url{#2}}

\bibitem[{\citenamefont{Trubeckov and Hramov}(2004)}]{Trubeckov}
\bibinfo{author}{\bibfnamefont{D.~I.} \bibnamefont{Trubeckov}}
  \bibnamefont{and} \bibinfo{author}{\bibfnamefont{A.~E.}
  \bibnamefont{Hramov}}, \emph{\bibinfo{title}{Lectures on microwave
  electronics for physicists}} (\bibinfo{publisher}{Moscow, Fizmatlit},
  \bibinfo{year}{2004}).

\bibitem[{\citenamefont{Silin and Sazonov}(1971)}]{Silin1971}
\bibinfo{author}{\bibfnamefont{R.~A.} \bibnamefont{Silin}} \bibnamefont{and}
  \bibinfo{author}{\bibfnamefont{V.~P.} \bibnamefont{Sazonov}},
  \emph{\bibinfo{title}{Slow Wave Structures}} (\bibinfo{publisher}{National
  Lending Library for Science and Technology, Boston SPA, England},
  \bibinfo{year}{1971}).

\bibitem[{\citenamefont{Gover and Livni}(1978)}]{Gover}
\bibinfo{author}{\bibfnamefont{A.}~\bibnamefont{Gover}} \bibnamefont{and}
  \bibinfo{author}{\bibfnamefont{Z.}~\bibnamefont{Livni}},
  \bibinfo{journal}{Optics Communications} \textbf{\bibinfo{volume}{26}},
  \bibinfo{pages}{375} (\bibinfo{year}{1978}).

\bibitem[{\citenamefont{Baryshevsky and Feranchuk}(1984)}]{Baryshevsky1984}
\bibinfo{author}{\bibfnamefont{V.}~\bibnamefont{Baryshevsky}} \bibnamefont{and}
  \bibinfo{author}{\bibfnamefont{I.}~\bibnamefont{Feranchuk}},
  \bibinfo{journal}{Phys. Lett. A} \textbf{\bibinfo{volume}{102}},
  \bibinfo{pages}{141} (\bibinfo{year}{1984}).

\bibitem[{\citenamefont{Baryshevsky}(1988)}]{Baryshevsky1988}
\bibinfo{author}{\bibfnamefont{V.}~\bibnamefont{Baryshevsky}},
  \bibinfo{journal}{Doklady Akademy of Science USSR}
  \textbf{\bibinfo{volume}{299}}, \bibinfo{pages}{6} (\bibinfo{year}{1988}).

\bibitem[{\citenamefont{Baryshevsky}(2000)}]{Baryshevsky1}
\bibinfo{author}{\bibfnamefont{V.}~\bibnamefont{Baryshevsky}},
  \bibinfo{journal}{NIM A} \textbf{\bibinfo{volume}{445}}, \bibinfo{pages}{281}
  (\bibinfo{year}{2000}), \eprint{LANL e-print archive physics/9806039}.

\bibitem[{\citenamefont{Baryshevsky et~al.}(2002)\citenamefont{Baryshevsky,
  Batrakov, Gurinovich, Ilienko, Lobko, Moroz, Sofronov, and
  Stolyarsky}}]{Baryshevsky2002}
\bibinfo{author}{\bibfnamefont{V.}~\bibnamefont{Baryshevsky}},
  \bibinfo{author}{\bibfnamefont{K.}~\bibnamefont{Batrakov}},
  \bibinfo{author}{\bibfnamefont{A.}~\bibnamefont{Gurinovich}},
  \bibinfo{author}{\bibfnamefont{I.}~\bibnamefont{Ilienko}},
  \bibinfo{author}{\bibfnamefont{A.}~\bibnamefont{Lobko}},
  \bibinfo{author}{\bibfnamefont{V.}~\bibnamefont{Moroz}},
  \bibinfo{author}{\bibfnamefont{P.}~\bibnamefont{Sofronov}}, \bibnamefont{and}
  \bibinfo{author}{\bibfnamefont{V.}~\bibnamefont{Stolyarsky}},
  \bibinfo{journal}{NIM A} \textbf{\bibinfo{volume}{483}}, \bibinfo{pages}{21}
  (\bibinfo{year}{2002}).

\bibitem[{\citenamefont{Baryshevsky and Gurinovich}(2011)}]{Baryshevsky2010}
\bibinfo{author}{\bibfnamefont{V.~G.} \bibnamefont{Baryshevsky}}
  \bibnamefont{and} \bibinfo{author}{\bibfnamefont{A.~A.}
  \bibnamefont{Gurinovich}} (\bibinfo{year}{2011}), \eprint{LANL e-print
  archive physics/1011.2983v1}.

\bibitem[{\citenamefont{Landau et~al.}(1984)\citenamefont{Landau, Lifshitz, and
  Pitaevskii}}]{Landau}
\bibinfo{author}{\bibfnamefont{L.}~\bibnamefont{Landau}},
  \bibinfo{author}{\bibfnamefont{E.}~\bibnamefont{Lifshitz}}, \bibnamefont{and}
  \bibinfo{author}{\bibfnamefont{L.}~\bibnamefont{Pitaevskii}},
  \emph{\bibinfo{title}{Electrodynamics of Continuous Media}}
  (\bibinfo{publisher}{Butterworth Heinemann}, \bibinfo{year}{1984}),
  \bibinfo{edition}{2nd} ed.

\bibitem[{\citenamefont{Jackson}(1998)}]{Jackson1965}
\bibinfo{author}{\bibfnamefont{J.}~\bibnamefont{Jackson}},
  \emph{\bibinfo{title}{Classical Electrodynamics}}
  (\bibinfo{publisher}{Wiley}, \bibinfo{year}{1998}), \bibinfo{edition}{3rd}
  ed.

\bibitem[{\citenamefont{Morse and Feshbach}(1953)}]{Mors1953}
\bibinfo{author}{\bibfnamefont{P.~M.} \bibnamefont{Morse}} \bibnamefont{and}
  \bibinfo{author}{\bibfnamefont{H.}~\bibnamefont{Feshbach}},
  \emph{\bibinfo{title}{Methods of Theoretical Physics}}
  (\bibinfo{publisher}{Mc Graw Hill, New York}, \bibinfo{year}{1953}).

\bibitem[{\citenamefont{Peter and Garate}(1992)}]{Peter1992}
\bibinfo{author}{\bibfnamefont{W.}~\bibnamefont{Peter}} \bibnamefont{and}
  \bibinfo{author}{\bibfnamefont{E.}~\bibnamefont{Garate}},
  \bibinfo{journal}{Phys. Rev. A} \textbf{\bibinfo{volume}{45}},
  \bibinfo{pages}{8833} (\bibinfo{year}{1992}).

\end{thebibliography}

\end{document}